\documentclass[aps,pra,reprint,amssymb,amsmath,superscriptaddress,noeprint]{revtex4-2} 
\usepackage[T1]{fontenc}
\usepackage[utf8]{inputenc}
\usepackage{graphicx}% Include figure files
\usepackage{dcolumn}% Align table columns on decimal point
\usepackage{bm}% bold math
\usepackage[colorlinks=true,citecolor=blue,linkcolor=blue,urlcolor=blue]{hyperref}
\usepackage{orcidlink}
\usepackage{physics}

\newcommand{\mean}[1]{\langle #1 \rangle}
\usepackage[
    ]{draftfigure}

\usepackage{titlesec}
\usepackage{placeins}
\titleformat{\section}[block]{\bfseries\large}{\relax}{}{}[]
\titlespacing*{\section}{0pt}{12pt}{6pt}

\titleformat{\subsection}[runin]{\bfseries}{}{0pt}{}[.]
\titlespacing*{\subsection}{0pt}{10pt}{1em}

\begin{document}

\title{Hybrid quantum-classical analog simulation of two-dimensional Fermi-Hubbard models with neutral atoms} 

\author{Sergi Julià-Farré\,\orcidlink{0000-0003-4034-5786}}
\email{sergi.julia-farre@pasqal.com}
\let\comma,

\affiliation{PASQAL SAS, 24 rue Emile Baudot - 91120 Palaiseau,  Paris, France}

\author{Antoine Michel\,\orcidlink{0000-0002-2211-5066}}
\let\comma,
\affiliation{Electricité de France, EDF Recherche et Développement, Département Matériaux et Mécanique des Composants,
Les Renardières, F-77250 Moret sur Loing, France}

\author{Christophe Domain\,\orcidlink{0000-0002-9802-9818}}
\let\comma,
\affiliation{Electricité de France, EDF Recherche et Développement, Département Matériaux et Mécanique des Composants,
Les Renardières, F-77250 Moret sur Loing, France}

\author{Joseph Mikael\, \orcidlink{0000-0002-0416-5297}}
\let\comma,
\affiliation{EDF R\&D, 7, boulevard Gaspard Monge, 91120 Palaiseau, France}

\author{Jacques-Charles Lafoucriere}
\let\comma,
\affiliation{CEA DAM, 91680, Bruyères-le-Châtel, France}

\author{Joseph Vovrosh\,\orcidlink{0000-0002-1799-2830}}
\let\comma,
\affiliation{PASQAL SAS, 24 rue Emile Baudot - 91120 Palaiseau,  Paris, France}

\author{Ahmed Chahlaoui\,\orcidlink{0009-0005-8113-7344}}
\let\comma,
\affiliation{PASQAL SAS, 24 rue Emile Baudot - 91120 Palaiseau,  Paris, France}

\author{Dorian Claveau}
%\email{sergi.julia-farre@pasqal.com}
\let\comma,
% \thanks{These two authors contributed equally to this work.}
\affiliation{PASQAL SAS, 24 rue Emile Baudot - 91120 Palaiseau,  Paris, France}

\author{Guillaume Villaret\,\orcidlink{0000-0002-3898-8646}}
%\email{sergi.julia-farre@pasqal.com}
\let\comma,
% \thanks{These two authors contributed equally to this work.}
\affiliation{PASQAL SAS, 24 rue Emile Baudot - 91120 Palaiseau,  Paris, France}

\author{Julius de Hond\,\orcidlink{0000-0003-2217-934X}}
%\email{sergi.julia-farre@pasqal.com}
\let\comma,
% \thanks{These two authors contributed equally to this work.}
\affiliation{PASQAL SAS, 24 rue Emile Baudot - 91120 Palaiseau,  Paris, France}

\author{Loïc Henriet}
\let\comma,
\affiliation{PASQAL SAS, 24 rue Emile Baudot - 91120 Palaiseau,  Paris, France}

\author{Antoine Browaeys\,\orcidlink{0000-0001-9941-8869}}
\let\comma,
\affiliation{Université Paris-Saclay, Institut d’Optique Graduate School,
CNRS, Laboratoire Charles Fabry, 91127 Palaiseau Cedex, France}
\author{Thomas Ayral\,\orcidlink{0000-0001-9941-8869}}
\let\comma,
\affiliation{Eviden Quantum Lab, 78340 Les Clayes-sous-Bois, France}
\author{Alexandre Dauphin\,\orcidlink{0000-0003-4996-2561}}
\email{alexandre.dauphin@pasqal.com}
\let\comma,
\affiliation{PASQAL SAS, 24 rue Emile Baudot - 91120 Palaiseau,  Paris, France}

\begin{abstract}
We experimentally study the two-dimensional Fermi-Hubbard model using a Rydberg-based quantum processing unit in the analog mode. Our approach avoids encoding directly the original fermions into qubits and instead relies on reformulating the original model onto a system of fermions coupled to spins and then decoupling them in a self-consistent manner. We then introduce the auxiliary spin solver: this hybrid quantum-classical algorithm handles a free-fermion problem, which can be solved efficiently with a few classical resources, and an interacting spin problem, which can be naturally encoded in the analog quantum computer. This algorithm can be used to study both the equilibrium Mott transition as well as non-equilibrium properties of the original Fermi-Hubbard model, highlighting the potential of quantum-classical hybrid approaches to study strongly correlated matter. 
\end{abstract}

\maketitle

The last decade has witnessed remarkable experimental progress in Rydberg-based quantum processing units (QPU)~\cite{browaeys_many-body_2020, saffman_quantum_2010, henriet_quantum_2020} of neutral atoms in programmable optical tweezers. Those QPUs can both be used as universal digital quantum computers, with recent progress towards quantum error correction~\cite{Bluvstein_2023, saffman_quantum_2025}, as well as analog quantum simulators~\cite{feynman_simulating_1982, daley_twenty-five_2023}. In both schemes, these platforms have the advantage of high flexibility in the layout geometry and scalability in the number of qubits~\cite{pichard_rearrangement_2024,manetsch_tweezer_2024}. 

In the analog mode, QPUs have been used in a quantum-classical workflow to address, among others, quantum simulation tasks or graph problems in optimization~\cite{ebadi_quantum_2022} and machine learning~\cite{albrecht_quantum_2023}. Here, QPUs serve as specialized computing units to tackle problems naturally embedded in the QPU. Classical computers then perform either pre- or post-processing, or the optimization of the parameters sent to the QPU.

Handling non-native problems poses greater challenges for the analog qubit platform, a paradigmatic case being the encoding of electronic models due to the nonlocality of fermion-to-qubit mappings.
A notable example is the Fermi-Hubbard model (FHM)~\cite{hubbard_electron_1997, hubbard_electron_1997-1, hubbard_electron_1997-2}, one of the biggest challenges of modern condensed-matter physics~\cite{arovas_hubbard_2022}.

In fact, solving this model has served as a standard benchmark to test the progress of the state-of-the-art numerical~\cite{qin_hubbard_2022} and experimental techniques over the last decades. From a quantum simulation perspective, fermionic ultracold atoms in optical lattices~\cite{bloch_many-body_2008, lewenstein_ultracold_2012, mazurenko_cold-atom_2017,shao_observation_2024} have historically represented a fruitful analog simulator of the model, recently achieving remarkably low temperatures~\cite{xu_neutral-atom_2025, Chalopin2024, kendrick_pseudogap_2025}. In digital quantum computers, despite the fundamental challenge of encoding a fermionic problem with qubits, recent works have shown progress towards a universal quantum simulation of the model either directly ~\cite{nigmatullin_experimental_2025, evered_probing_2025} or through embedding methods \cite{Ayral2025b}.

\begin{figure*}[t]
\centering
\includegraphics[]{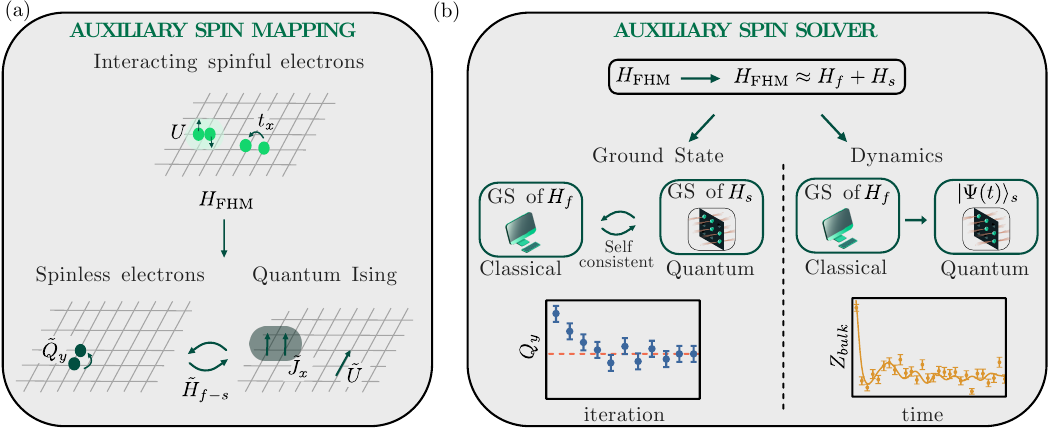}
\caption{(a) Sketch of the anisotropic Fermi-Hubbard model and the auxiliary spin mapping: the original interacting spinful Hamiltonian is mapped to a model of spinless fermions coupled to a quantum Ising model. (b) Hybrid quantum-classical solver. A mean-field decoupling is performed on $\tilde{H}_{f-s}$. Then the solver can be used to either solve the ground-state physics through a hybrid quantum-classical loop or to solve the dynamics with the help of the QPU.}
\label{fig:intro}
\end{figure*}

In this work, we show the capabilities of qubit-based analog, i.e., non-universal, QPUs to investigate the physics of the two-dimensional (2D) FHM via a recently proposed classical-quantum hybrid algorithm~\cite{michel_hubbard_2024}. This approach, sketched in Fig.~\ref{fig:intro}, circumvents the fermion-to-qubit mapping problem by means of a \emph{$Z_2$ slave spin} reformulation~\cite{ruegg_z_2010, hassan_slave_2010}. This leads, after a mean-field decoupling, to a description in terms of noninteracting fermionic degrees of freedom, classically solvable in any regime, and interacting spin degrees of freedom, which can be naturally encoded in the Rydberg-based analog QPU. In other words, we can perform fermionic many-body calculations using spin degrees of freedom. In the algorithm, the classical fermionic solver and the QPU are used iteratively in a self-consistent workflow. Compared to the original $Z_2$ slave spin method~\cite{ruegg_z_2010, hassan_slave_2010}, where the auxiliary spins are handled within a single-site or few-site cluster mean-field approximation, the direct solution of the spin problem by means of the analog QPU allows for an exact treatment of the strong spin-spin correlations in a larger unit cell. Moreover, this scheme can take advantage of the layout flexibility of programmable tweezer arrays to simulate FHMs with different lattice geometries. 

To demonstrate the feasibility and utility of this approach, we experimentally realize the proposed hybrid quantum-classical algorithm to study a 2D FHM on our Orion Alpha QPU accessible on the cloud. In this work we restrict ourselves to 36 Fermi-Hubbard sites, corresponding to 36 atoms. This choice reflects a balance between computational time and the ability to benchmark against classical calculations. While this constitutes a relatively modest problem size, it is large enough to demonstrate the method, and small enough to enable quantitative validation. Importantly, the approach is not limited to this scale: already on existing hardware one could reach $\gtrsim 200$ sites~\cite{scholl_quantum_2021, ebadi_quantum_2021}, and in the longer term explore thousands in the thermodynamic limit~\cite{pichard_rearrangement_2024, manetsch_tweezer_2024}, albeit without classical validation.

On the one hand, we study the equilibrium Mott transition in an anisotropic version of the model using the 36-qubit Rydberg simulator in a rectangular lattice.  On the other hand, we also consider the nonequilibrium properties of the isotropic model in the square lattice after a sudden quench of the on-site interaction, requiring a 36-qubit Rydberg simulator in a square lattice. We observe the expected collapsed oscillations of the quasiparticle weight in the Mott phase~\cite{schiro_quantum_2011, yang_benchmarking_2019, eckstein_thermalization_2009, iyer_coherent_2014, will_observation_2015, riegger_interaction_2015}. 
 
\begin{figure*}[t!]
  \centering
  \includegraphics[width=\textwidth]{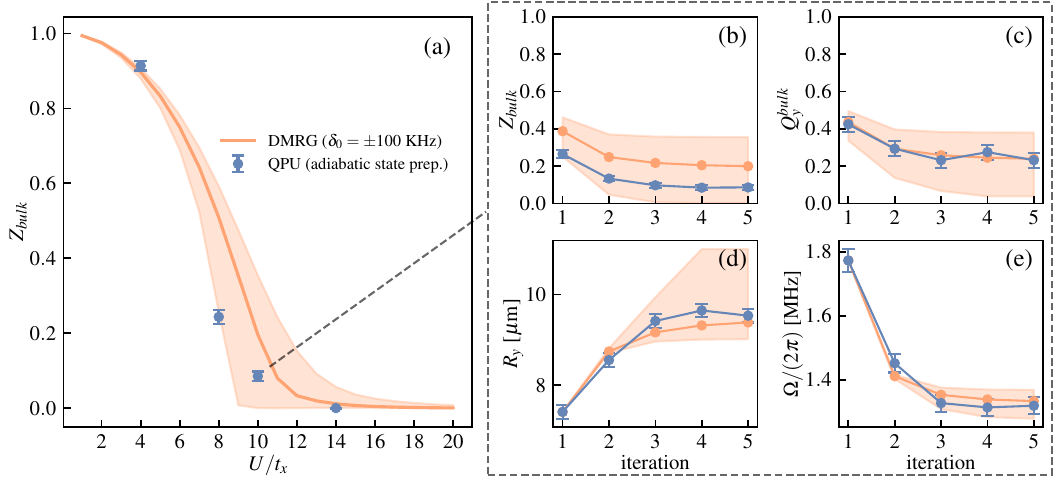}
  \caption{Results for the equilibrium loop. (a) Metal-to-Mott transition in the anisotropic FHM ($t_y=0.65t_x$) using the auxiliary spin mapping and the QPU or DMRG to solve the ground state of the spin problem. The uncertainty area of the DMRG line accounts for a systematic error in the QPU detuning, $\delta_0$.  (b)-(e) Variation of the observables and QPU setpoints through the iterative loop of the hybrid quantum-classical solver.}
  \label{fig:figure2}
\end{figure*}

\section*{Results}

\subsection*{Anisotropic Fermi-Hubbard model in the auxiliary spin approximation}
We consider the 2D FHM on the anisotropic square lattice
\begin{equation}
H_\text{FHM}=\sum_{\alpha=x,y}-t_{\alpha}\sum_{i,\sigma}(d^\dagger_{i,\sigma}d_{i+\alpha,\sigma}+\text{H.c.})+\frac{U}{2}\sum_i(n^d_i-1)^2,
\label{eq:fermihubbard}
\end{equation}
where $d_{i,\sigma}\,(d^\dagger_{i,\sigma})$ are the annihilation (creation) fermionic operators on the $i$-th site of a rectangular lattice of $N=L\times L$ sites with internal spin $\sigma \in \{\uparrow\,,\downarrow\}$, and $n^d_i$ the total occupation number at site $i$. Here $t_{\alpha}$ ($\alpha=x,y$) is the tunneling amplitude between nearest neighbors (NN) in the $x$ and $y$ direction. $U>0$ is the on-site Hubbard repulsive interaction, penalizing single-site double occupations. In this work, we also assume half-filling occupation $\langle n^d_i \rangle = 1$. This has been implicitly enforced in Eq.~\eqref{eq:fermihubbard} by setting the chemical potential to $\mu=U/2$. This model is in general hard to solve at $T=0$ away from the perturbative limits $U/t_\alpha\ll 1$ and $U/t_\alpha\gg 1$~\cite{leblanc_solutions_2015}. From an equilibrium perspective, the ground-state phase diagram of the isotropic case ($t_\alpha=t$) at half-filling is qualitatively understood~\cite{schafer_tracking_2021, schafer_fate_2015, simkovic_extended_2020, kim_spin_2020}. The free-electron metallic state at $U=0$ evolves towards a strongly-correlated insulator at large $U/t$ through the so-called \emph{metal-to-Mott transition}. For the anisotropic case ($t_x\neq t_y$), the results are scarce, but point towards a similar picture away from the quasi-1D limit~\cite{moukouri_mott_2011,moukouri_universality_2012,perez-conde_two-dimensional_1992}. In contrast, few works address the dynamical properties of the model, and only in  specific regimes~\cite{schiro_quantum_2011, yang_benchmarking_2019, eckstein_thermalization_2009, iyer_coherent_2014, will_observation_2015, riegger_interaction_2015}. 

We map the original model to an extended Hilbert space of fermions and spins, where the FHM is expressed as $H_\text{FHM}=\tilde{H}_{f-s}$ (see Methods and Ref.~\cite{ruegg_z_2010}). After a mean-field decoupling of $\tilde{H}_{f-s}$, the Hamiltonian can be written as a sum of two self-consistent terms, i.e.,  $H_\text{FHM}\approx H_\text{f}+H_{s}$, where
\begin{equation}\label{eq:auxiliaryfermion}
H_\text{f}=-\sum_{i} \sum_{\alpha=x,y} Q_{\alpha}(f^\dagger_{i}f_{i+\alpha}+\text{H.c.}),
        \end{equation}
\begin{equation}\label{eq:auxiliaryspin}
H_\text{s}=-\sum_{i}\sum_{\alpha=x,y} J_{\alpha}\sigma^z_i\sigma^z_{i+\alpha} +\frac{U}{4}\sum_i\sigma^x_i.  
\end{equation}
Here $Q_{\alpha}=t_\alpha\mean{\sigma^z_i\sigma^z_{i+\alpha}}$ is the NN fermion hopping in the $\alpha=x,y$ direction, and $J_{\alpha}=4t_\alpha\text{Re}\mean{f^\dagger_{i} f_{i+\alpha}}$ is the corresponding NN spin coupling. A central quantity of interest is the quasiparticle weight $Z$ of the original Fermi-Hubbard model, which can be expressed as the z-component of the auxiliary spin operator expectation value $Z=\frac{1}{N}|\sum_i \mean{\sigma^z_i}|^2$. An important remark here is that in Eqs.~\eqref{eq:auxiliaryfermion}-\eqref{eq:auxiliaryspin} we used translational invariant values for $Q_\alpha$ and $J_\alpha$. However, this approximation is only strictly valid in the thermodynamic limit or under periodic boundary conditions. Thus, to mitigate edge effects when solving the spin system in a finite system, we use the bulk-averaged quantities $Z^\text{bulk}$ ($Q^\text{bulk}_\alpha$), measured in the sites (bonds) of the center plaquette of the $L\times L$ square lattice.

While the free-fermion system described by Eq.~\eqref{eq:auxiliaryfermion} can be solved efficiently with a classical workstation, we use the Orion Alpha QPU, a Rydberg-based neutral-atom device, to find the ground state or to study the unitary dynamics of Eq.~\eqref{eq:auxiliaryspin}. In particular, we use $L\times L=6\times 6$ rectangular registers characterized by the NN separation in $x$ and $y$, namely $R_x$ and $R_y$, which realize the following Hamiltonian~\cite{browaeys_many-body_2020}
\begin{equation}\label{eq:ham_rydberg_main}
H^{\text{QPU}}_\text{s}\approx \sum_{i}\sum_{\alpha=x,y} \frac{C_6}{4R_\alpha^6}\sigma^z_i\sigma^z_{i+\alpha} -\frac{\hbar\Omega(t)}{2}\sum_i\sigma^x_i\equiv -H_s.  
\end{equation}
Here, the local qubit is encoded in a state of the $^{87}$Rb ground-state manifold $\ket{0} = |5S_{1/2}, F=2,m_F=2\rangle$ and a Rydberg state $\ket{1} = |60S_{1/2}, m_J = 1/2\rangle$. $C_6>0$ is the strength of the van der Waals interaction due to the off-resonant dipole-dipole Rydberg interaction. $\Omega(t)$ is the effective single-photon Rabi frequency addressing the transition. An important remark is that the Rydberg Hamiltonian $H^\text{QPU}_\text{s}$ is usually expressed in terms of density-density interactions $n_in_j$, with $n_i\equiv (1+\sigma^z_i)/2$, and a laser detuning term $-\hbar \delta(t)/2\sum_in_i$. Here, to map the spin Hamiltonian of interest, namely $H_\text{s}$, we set the laser detuning to a value $\delta(t)=\delta (R_x, R_y)$ that minimizes the presence of linear terms in $\sigma^z_i$ when going from the density $n_i$ to the z-spin $\sigma^z_i$ operator representation (see details in the Methods section).
Finally, note that, due to time-reversal symmetry, the relative sign between $H^{\text{QPU}}_\text{s}$ and $H_\text{s}$ is physically irrelevant for the unitary dynamics under consideration~\cite{frerot_multi-speed_2018} .

\subsection*{Hybrid simulation of the equilibrium phase diagram}

We start by finding the equilibrium state of the FHM at a fixed $t_x/t_y=0.65$. To this aim, we compute the fermionic ground state of Eq.~\eqref{eq:auxiliaryfermion} with a classical workstation, and the spin ground state of Eq.~\eqref{eq:auxiliaryspin} with the Orion Alpha QPU, consisting of an array of $6\times 6$ atoms. For the spin groundstate, we rely on a quasi-adiabatic state preparation protocol~\cite{browaeys_many-body_2020} (see details in the Methods). At each value of $U/t_x$, we start from an initial guess for the fermionic expectation value $J_\alpha$, which defines the initial setpoint of the QPU as $J_\alpha/U= C_6/(8\hbar\Omega R_\alpha^6)$. Subsequently, we iteratively solve the two coupled systems until achieving convergence of the quasiparticle weight $Z^\text{bulk}$. In the QPU, the change of the spin Hamiltonian between iteration steps implies a change in the control parameters. Specifically, we fix $R_x$ and vary $R_y$ and $\Omega$ throughout the loop. This is achieved within an autonomous workflow involving both the workstation and the cloud-accessed QPU.

The results are shown in Fig.~\ref{fig:figure2}. The hybrid classical-quantum solver reveals the metal-to-Mott transition, with the expected monotonous decay of the quasiparticle weight $Z^\text{bulk}$ for increasing interaction $U$, as can be seen in Fig.~\ref{fig:figure2}(a). Moreover, we test the accuracy of our QPU annealing solver by replacing it with a ground-state numerical solver of $H^\text{QPU}_s$ in the hybrid algorithm, shown as the orange solid line in Fig.~\ref{fig:figure2}(a). The numerical solver is based on the density-matrix-renormalization group (DMRG) method for matrix-product-states (MPS), and is very accurate for this $6\times 6$ spin system. We observe that the QPU results are compatible with the ground-state in the presence of expected QPU residual offsets (orange shaded area). Such orange uncertainty area also highlights the role of experimental uncertainties in the QPU results; reducing this uncertainty is a natural perspective for future hardware improvements. For completeness, in the Supplemental Materials we also show, numerically, the effect of the approximate mapping of the Ising model $H_s$ in the implemented $H^\text{QPU}_s$. While for small systems edge effects lead to sizable differences, at sufficiently large system sizes the difference in the equilibrium curves are only due to the fast $1/r^6$ decay of interactions beyond NN in $H^\text{QPU}_s$.

It is also instructive to consider the evolution of the QPU setpoint and measured observables during the iterative classical-quantum loop. This can be seen in Figs.~\ref{fig:figure2}(b)-(e), for the datapoint $U/t_x=10$. On the one hand, in Figs.~\ref{fig:figure2}(b)-(c), the physical observables of the QPU converge to their equilibrium value, in agreement with the DMRG simulations of the ground-state. On the other hand, Figs.~\ref{fig:figure2}(d)-(e) show how the $R_y$ and $\Omega$ must be changed at each step of the algorithm to simulate the required Ising model. We stress that a very precise calibration of the atomic positions and laser parameters is crucial for the accuracy of the algorithm, especially close to the transition (see Methods). In terms of runtime, each step of the iterative loop consists of $500$ pre-calibration QPU shots and $500$ shots to estimate the local magnetization and NN correlations of the spin ground state, leading to $1000$ shots for each annealing simulation. Since we use 5 iterations to ensure the convergence of the algorithm, we use $5000$ shots for each phase diagram data point in Fig.~\ref{fig:figure2}, which take $\approx 2$ hours at a repetition rate of $\approx 0.65$ Hz. Looking ahead, an increase of the repetition rate would proportionally reduce the runtime. In contrast, finding the free-fermion equilibrium state takes a few seconds in a classical workstation.  

\subsection*{Hybrid simulation of post-quench dynamics}
We consider the dynamics occurring after a sudden quench of the Hubbard interaction $U_f$ in the metallic state of $H_\text{FHM}$ present at $U=0$. Since even for the isotropic case the results are scarce, here we focus on this model with $t_x=t_y=t$. However, note that the methods described below can be equally applied to the dynamics of anisotropic models. 

Within the auxiliary spin mapping, the interaction quench dynamics present a major simplification~\cite{michel_hubbard_2024}: the eigenstates of $H_\text{f}$ are always Bloch states independently of the values of $Q_{x(y)}$, and thus the initial fermionic state at $U=0$ is also an eigenstate at $U=U_f$. This leads to trivial dynamics in the auxiliary fermionic subsystem under the auxiliary-spin mapping and the assumption of lattice translational invariance. That is, we only need the fermionic expectation values $J_{x(y)}$ in the $U=0$ equilibrium state to simulate the dynamics of the auxiliary spin system. We are then left with the problem of solving the dynamics of the interacting spin problem
\begin{equation}
    \ket{\Psi(t)}_\text{s}=e^{-itH_\text{s}}\ket{0}^{\otimes N}_\text{s},
\end{equation}
which is a much harder computational task than the equilibrium case discussed in the previous section. 

\begin{figure}[t!]
    \centering
    \includegraphics[width=1\linewidth]{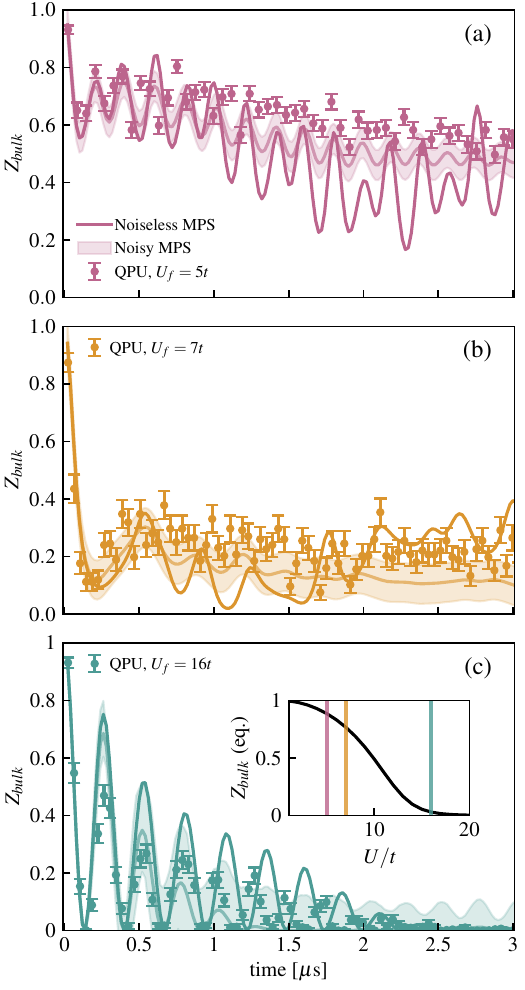}
    \caption{Dynamics of the quasiparticle weight $Z_\text{bulk}$ after a sudden interaction quench in a square lattices of size $N=6\times 6$. We compare the QPU data with ideal MPS simulations of the QPU Hamiltonian, and also MPS simulations of the noise model. The inset in (c) shows the equilibrium curve of $Z_\text{bulk}$ for the the same $6\times 6$ square lattice, using the purely classical auxiliary-spin method. The values of post-quench $U_f$ in (a)-(c) are indicated as vertical lines.}
    \label{fig:figure3}
\end{figure}

Here we simulate such spin post-quench dynamics with the Orion Alpha QPU for a $N=6\times 6$ system for three different values of the quenched interaction $U_f$. For each of them, we monitor the many-body spin dynamics by measuring the QPU state at different final times. We furthermore compare the QPU results with noiseless MPS dynamics and noisy simulations of the spatial noise in the registers and the laser phase noise.

Figure~\ref{fig:figure3} shows the behavior of the quasiparticle weight in the time domain for the different scenarios. In the first place, Fig.~\ref{fig:figure3}(a) shows the strong spin interaction regime, with $U_f \approx 5t$, corresponding to the weak interaction regime in the FHM. As expected, the quasiparticle weight stays at a finite value, signaling this quench to be within the metallic phase. Moreover, we observe that the numerical noise model accurately predicts the damping of the oscillations, which exhibit a plateau at a finite $Z_\text{bulk}$. We also provide the Fourier-transformed signal in the supplementary Fig.~\ref{fig:figureS2}(a). One prediction of this metallic regime is that the quasiparticle weight exhibits a frequency mode at $\omega/(2\pi)\approx U_f/2$, which however is masked here even in the noiseless numerical simulation, due to finite-size effects and a limited evolution time. 

If we further increase the interaction to $U_f\approx 7t$, we approach the metal-Mott transition from the metallic side. As can be seen in Figure~\ref{fig:figure3}(b), while the quasiparticle weight stays finite during the post-quench dynamics, it also approaches a zero value. Similarly to the previous case, the noise model captures the discrepancy between the QPU and exact dynamics. In the frequency domain, see supplementary Fig.~\ref{fig:figureS2}(b), a clean mode picture is less clear in this regime, even in the large time and system size limits, since many different modes are expected to play a role in such a crossover regime between post-quench dynamics in the metallic and Mott regions. In accordance, we do not observe any clear peak in the QPU signal, which is furthermoe impacted by noise.

In the Mott region, at relatively large interactions, the Fermi-Hubbard model is expected to exhibit collapsed oscillations of the quasiparticle weight with revivals at latter times. That is, the quasiparticle nature of excitations is expected to vanish after a transient oscillatory behavior at early times in the post-quench dynamics and eventually increase again. Figure~\ref{fig:figure3}(c) shows such a behavior in the Mott phase with a large final interaction $U_f=16t$. The QPU simulation captures the first collapse regime of the noiseless MPS simulation, with even better agreement when comparing with the noisy MPS simulation. In this case, the frequency analysis of the QPU data reveals that the early-time oscillations occur at a frequency $\omega/(2\pi)\approx U_f$, as observed in numerical simulations and theoretical predictions. Finally, it is worth mentioning that a clear prospect would be to increase the coherence time of the device to be able to observe the revival of oscillations.

\section*{Discussion}
In this work, we experimentally implemented a hybrid quantum-classical algorithm based on the auxiliary-spin method for the 2D FHM. In this algorithm, a Rydberg-based QPU performed the analog simulation of the parent spin Hamiltonian, whereas the parent fermionic problem was solved in a classical workstation requiring few computational resources. We have shown that the algorithm can be used both to study equilibrium and nonequilibrium properties of the original FHM, the main difference being that the equilibrium algorithm requires automating the acquisition of data to feed the classical computer and realizing a coordinated self-consistent loop of the method. 

On the one hand, we used the algorithm to study ground-state properties of an anisotropic FHM consisting of 36 sites, finding the expected metallic-Mott transition. On the other hand, we studied the out-of-equilibrium dynamical response of a 36-site isotropic FHM. We observed that the quasiparticle weight remains finite for interaction quenches within the metallic phase, and exhibits collapsed oscillations when the interaction quench drives the system into the Mott regime. Both in the equilibrium and nonequilibrium studies, we found qualitative agreement between the QPU simulations and exact numerical emulations of the auxiliary-spin method, and we found that the quantitative discrepancies can be explained by a QPU noise model that accounts for Hamiltonian parameter offsets, position noise, and laser phase noise.

While here we focused on a particular example of the usage of the algorithm at equilibrium and nonequilibrium, an important remark is that the procedure presented in this work could be generalized to many other scenarios. In terms of lattice geometries, the auxiliary-spin mapping can also be easily applied to other bipartite lattices beyond rectangular ones, such as e.g., the honeycomb geometry. For non-bipartite lattices (such as the triangular), the method could also be applied with an extra step in the classical-quantum workflow to fix the particle filling (see, for instance, Refs.~\cite{ruegg_z_2010, yang_benchmarking_2019, medici_modeling_2017, hassan_slave_2010}). Such an extra step would also open the possibility of studying the model with an arbitrary electron filling. In terms of transition regimes, besides the metallic-Mott transition driven by the Hubbard interaction $U$, it would also be interesting to study scans over $t_y/t_x$, e.g., to study the 1D to 2D crossover. Finally, in terms of system sizes, the scalability of the algorithm is linked to the ability of the platform to increase the number of atoms while maintaining the coherence. Since recent works reported the loading of thousands of neutral atoms in tweezer arrays~\cite{manetsch_tweezer_2024,pichard_rearrangement_2024}, implementing the auxiliary-spin method close to the thermodynamic limit is also a clear prospect for the near future. In such a limit, the numerical accuracy of state-of-the-art numerics, such as the MPS ansatz used in this work, is expected to worsen dramatically due to the exponential increase of the Hilbert space. 

\section*{Methods}

\subsection*{Auxiliary-spin mapping} 

Here we outline the $\mathbf{Z}_2$ auxiliary spin representation of the Fermi-Hubbard model~\cite{ruegg_z_2010, michel_hubbard_2024}. One represents the original spinful fermionic operator at each lattice site as $d^\dagger_{i,\sigma}=\sigma^z_if^\dagger_{i,\sigma}$, where $f^\dagger_{i,\sigma}$ is a pseudo-fermion operator and $\sigma^z_i$ the Pauli operator of the auxiliary spin. Note that in this representation the local Hilbert space at each lattice site is enlarged and one needs to impose the local site constraint $\sigma^x_i+1=2(n^f_i-1)^2$. In the case under consideration, a bipartite square or rectangular lattice at half filling, this constraint is automatically fulfilled~\cite{schiro_quantum_2011}. When written in the spin-fermion representation, the FHM of Eq.~\eqref{eq:fermihubbard} reads
\begin{equation}
    \tilde{H}_{f-s}=-\sum_{\alpha=x,y}t_\alpha\sum_{i,\sigma}(\sigma^z_i\sigma^z_{i+\alpha}f^\dagger_{i,\sigma}f_{i+\alpha,\sigma}+\text{H.c.})+\frac{U}{4}\sum_i\sigma^x_i.
\end{equation}
where we have used the constraint $\sigma^x_i+1=2(n^f_i-1)^2$, the half-filling condition $\mu=U/2$, and dropped a physically irrelevant constant.

To arrive to the self-consistent auxiliary spin equations, we perform a mean-field decoupling of the spin-fermion interaction $\sigma^z_i\sigma^z_jf^\dagger_{i,\sigma}f_{j,\sigma}\approx \langle \sigma^z_i\sigma^z_j\rangle f^\dagger_{i,\sigma}f_{j,\sigma}+\sigma^z_i\sigma^z_j\langle f^\dagger_{i,\sigma}f_{j,\sigma}\rangle-\langle\sigma^z_i\sigma^z_j\rangle \langle f^\dagger_{i,\sigma}f_{j,\sigma}\rangle$. Moreover, due to spin-species symmetry of the resulting self-consistent mean-field theory, we work with spinless fermions $f^\dagger_{i,\sigma}\to f^\dagger_{i}$. To account for the two-species nature of the problem, we simply add a factor of 2 to express spinful expectation values from spinless ones, e.g., $\sum_\sigma\langle f^\dagger_{i,\sigma}f_{j,\sigma}\rangle = 2\langle f^\dagger_{i}f_{j}\rangle$. Finally, to reduce the number of self-consistent mean-field parameters that the algorithm needs to determine, we assume that they are translationally invariant. That is, we consider that $\langle f^\dagger_{i}f_{j}\rangle$ and $\langle \sigma^z_{i}\sigma^z_{j}\rangle$ can only depend on the orientation of the bond $(i,j)$ (along $x$ or $y$) but not on the site indexes. Within this approximation, which is exact in the thermodynamic limit or under periodic boundary conditions, one arrives to Eqs.~\eqref{eq:auxiliaryfermion}-\eqref{eq:auxiliaryspin} of the main text. 

\subsection*{Free-fermion solution} 

The free-fermionic Hamiltonian \eqref{eq:auxiliaryfermion} can  readily be solved with the help of the Bloch's theorem. The analytical and semi-analytical formulas here below thus allow one to efficiently solve the fermionic part of the problem at zero temperature and for large $N$. In the reciprocal space, the fermionic Hamiltonian exhibits a band structure of the form
\begin{equation}\label{eq:fermionsolution1}
E_\mathbf{k}=-2Q_x\cos(k_x)-2Q_y\cos(k_y),
\end{equation}
where $k_\alpha \in [0, 2\pi)$ with a discrete step $\Delta k_\alpha =2\pi/N_\alpha$. Furthermore, $J_{\alpha}$ can be expressed as
\begin{equation}\label{eq:fermionsolution2}
    J_{\alpha} =\frac{2t_\alpha}{L}\sum_\mathbf{k} \cos(k_\alpha)(1+e^{\beta E_\mathbf{k}})^{-1},
\end{equation}
with $\mathbf{k}=(k_x,k_y)$, and $\beta=1/(k_B T)$ being the inverse temperature, which is set to $\beta\to \infty$ to study ground-state properties. In a finite-system, however, the temperature needs to be fixed at a small but finite value to avoid finite-size gaps. We set $k_BT/t_x=0.05$ for a large system consisting of $400$ sites. Note also that, in Eq.~\eqref{eq:fermionsolution2}, the thermal exponent in the Fermi-Dirac distribution, $e^{\beta \left[E_\mathbf{k}-(\mu-U/2)\right]}$ has been written under the half-filling condition of the FHM, $\mu=U/2$. Since, as a consequence of the particle-hole symmetry, the free-fermion energies of Eq.~\eqref{eq:fermionsolution1} appear in pairs $\pm |E_\mathbf{k}|$, it is straightforward to see that, when $\beta \to\infty$ the half-filling condition is achieved by populating the states with negative energies.

\subsection*{Quantum Ising model in the QPU} 

Solving the many-body 2D spin Hamiltonian~\eqref{eq:auxiliaryspin} with classical numerical methods for large $N$ typically requires state-of-the-art numerical methods and large computational resources. In particular, tensor network or quantum Monte Carlo methods are powerful tools to study equilibrium properties of the system, while arbitrary dynamics can only be characterized for short times in certain parameter regimes.  As an alternative, here we propose an analog simulation scheme, as the model is particularly suited for arrays of Rydberg atoms.   The atoms can be arranged in arbitrary positions of the 2D plane thanks to the layout flexibility, and they realize the Hamiltonian
\begin{equation}\label{eq:hamiltonian_fresnel1}
    H_{\text{QPU}} = \sum_{i<j}\frac{C_6}{4r_{ij}^6}\sigma^z_i\sigma^z_j+
    \sum_i \frac{\hbar \Omega(t)}{2}\sigma^x_i -\left[\frac{\hbar \delta(t)}{2}-\tilde{J}(\mathbf{r}_i)\right]\sigma^z_i.
    \end{equation} 
Here the local qubit is encoded in a state of the $^{87}$Rb ground-state manifold $\ket{0} = |5S_{1/2}, F=2,m_F=2\rangle$ and a Rydberg state $\ket{1} = |60S_{1/2}, m_J = 1/2\rangle$. $C_6>0$ is the strength of the van der Waals interaction due to the off-resonant dipole-dipole Rydberg interaction and $r_{ij}$ is the distance between sites $i$ and $j$ in the 2D plane. $\Omega(t)$ and $\delta(t)$ are the Rabi frequency and detuning of the external laser field, $\tilde{J}(\mathbf{r}_i)\equiv\sum_{j\neq i}\frac{C_6}{4r_{ij}^6}$ is the interaction strength per site, and we neglected a constant energy shift. In order to simulate the spin Hamiltonian~\eqref{eq:auxiliaryspin} we can identify $\hbar\Omega=U/2$, $\hbar\delta=2J_0$, with the definition $J_0\equiv \tilde{J}(\mathbf{r}_0\in \text{bulk})$, and  $|J_{x(y)}|=C_6/(4R_{x(y)}^6)$, $R_{x(y)}$ being the distance between $x(y)$ nearest-neighbors in the tweezer array. Hence, the anisotropic spin-spin interaction is achieved via a rectangular layout with $R_x\neq R_y$. Note also that while the Hamiltonian of Eq.~\eqref{eq:auxiliaryspin} consists of ferromagnetic interactions when $J_{\alpha}> 0$ , we can still use a QPU with antiferromagnetic couplings $C_6>0$, since the evolution under $H_{\text{QPU}}$ and $-H_{\text{QPU}}$ are indistinguishable for initial states with real coefficients in the computational basis~\cite{frerot_multi-speed_2018}.
With this choice of QPU parameters, we can write the error Hamiltonian as
\begin{equation}\label{eq:errorham}
 H_{\text{QPU}}-H_\text{s}=-\sum_{\substack{i<j \\ j \not\in (i+x, i+y)}}\frac{C_6}{4r_{ij}^6}\sigma^z_i\sigma^z_j+\sum_{i} \left[\tilde{J}_0-\tilde{J}(\mathbf{r}_i)\right]\sigma^z_i.
\end{equation}
The first term accounts for the interactions beyond NN that are present in the QPU. While its effect is typically small due to the fast $r_{ij}^{-6}$ decay of Van der Waals interactions, and is not expected to change qualitatively the physics of the model, these long-range interactions in general shift the parameter regimes of the original model. The second term is only finite for $i$-th sites that are close to the edge of the system, where the average interaction is reduced. While this term does not affect the bulk physics for a sufficiently large system, it enhances edge effects in finite systems and it could be removed using local detunings $\hbar\delta_i=2\tilde{J}(\mathbf{r}_i)$.

\subsection{Hybrid groundstate solver via QPU cloud access}

The above-mentioned setup allows for the study of the groundstate of $H_{\text{FHM}}$ through the following workflow involving a workstation, which solves $H_f$, connected to the QPU, which solves $H_s$, via cloud access. 

    \textit{1. Fixing of the Fermi-Hubbard setpoint and energy scale}. At a given value $U/t_x$, we fix the NN separation $R_x$, which will remain constant during the loop and will define the relevant energy scale. In particular, in the equilibrium phase diagram of the main text Fig.~\ref{fig:figure2} we fixed $R_x[\mu\text{m}]=(6.5,\, 7.1,\, 6.9, 7.2)$ for, respectively, $U/t_x=(4, 8, 10, 14)$.
    
   \textit{2. Fermionic ansatz initialization.} We start with an ansatz for the fermionic observables $J_\alpha=J_\alpha^0$, which defines our initial spin Hamiltonian $H_s^0$. 
   
    \textit{3. Register and laser calibrations.} The workstation computes the required QPU setpoint $(R_x, R_y, \delta, \Omega)$, and sends two calibration jobs. The first one consists of a single shot and is just used to equalize the required rectangular layout defined by $(R_x, R_y)$. The second one, consisting of 500 shots, performs a Ramsey spectroscopy experiment that calibrates the laser setpoint ($\delta, \Omega$). Note that, while the laser setpoints are already calibrated in the cloud QPU, performing this extra calibration right before the simulation improves further the accuracy of such calibration.
    
    \textit{4. Ising solver}. The QPU finds the approximate groundstate of $H_s$ via a quasiadiabatic state preparation protocol for $H_\text{QPU}$ to the setpoint $(R_x, R_y, \delta, \Omega)$. The pulse sequence is parametrized simply by a linear ramp $\tilde{\Omega}(t)=t\Omega/t_\text{max}$ at a constant detuning $\delta$, and with $t_\text{max}=4\,\mu$s. Note that the atoms are initialized in the noninteracting state $\ket{0}^{\otimes N}$, which corresponds to the highest excited state of $H_\text{QPU}$ for $\Omega=0, \delta > 0$. A destructive measurement repeated $500$ times at a final time $t_\text{max}$, provides access to bitstrings in the z-basis. 
    
    \textit{5. Ising solver post-processing.} The workstation collects the bitstrings in the z-basis to compute $Q_\alpha$ and the quasiparticle weight $Z$, correcting for measurement errors (see Methods section below). Since we are interested in the translational invariant value of these quantities in the thermodynamic limit, we measure their spatial average over the center-most sites of the lattice, to mitigate the impact of finite-system edge effects on these local observables. 
    
    \textit{6. Fermionic solver.} The workstation uses the values $Q_\alpha$ to solve the fermionic Hamiltonian $H_f^0$ and obtains iterated values for the fermionic observables $J_\alpha^1$, which define a new spin Hamiltonian $H_s^1$.
    
    \textit{7.} The loop starts over from step 3, and is iterated 5 times to ensure convergence in $Q_\alpha^\text{bulk},\, J_\alpha$, and $Z^\text{bulk}$.
    
\subsection{Hybrid solver for quench dynamics via QPU cloud access}

Due to the stationary behavior of the fermions in the post-quench dynamics considered here, simulating dynamics with the hybrid algorithm does not require a self-consistent workflow between the QPU and a classical workstation. Hence, in this case, the workflow is simply:

\textit{1. Fix the initial fermionic state.} We compute the equilibrium values of the fermionic observables $J_\alpha$ in the trivial metallic state present at $U=0$, for the desired $t_x$ and $t_y$. In this work we use $t_x=t_y=t$.

\textit{2. Register and laser calibrations.} As in the equilibrium solver, the workstation computes the required QPU setpoint $(R_x, R_y, \delta, \Omega)$, and sends a calibration job to equalize the layout, and another one to calibrate the laser setpoint ($\delta, \Omega$). In particular, we used the setpoints $R[\mu\text{m}]=(6.2, 6.5, 7.5)$ and $\Omega [\text{rad}/\mu{\text{s}}]=(11.76, 12.4, 12.01)$ for the quenches at $U_f/t=(5, 7, 16)$ of Fig.~\ref{fig:figure3} of the main text.

\textit{3. Post-quench dynamics simulation.} The QPU is used to monitor the spin dynamics after a sudden quench to a fixed value of $\Omega$ and $\delta$ (a constant pulse). To this aim, we take between 50 and 75 final times between $24$ and $3000\,$ns, and for each of them we estimate the observables of interest using between $200$ and $300$ shots.

\subsection{Correction of measurement errors} In the measurement process there are unavoidable measurement errors to due false Rydberg state detection, with a probability $\epsilon \approx 0.01$, and false ground-state detection, with a probability $\epsilon '\approx 0.07$. Since these error probabilities are pre-calibrated, we can use them to correct the measured expectation values. In particular, if we denote the measured local magnetization as $\widetilde{\langle \sigma^z_i \rangle}$, then the corrected value can be written as:
\begin{equation}\label{eq:spam_magn}
\langle \sigma^z_i \rangle= \frac{\widetilde{\langle \sigma^z_i \rangle}+(\epsilon ' -\epsilon)}{1-\epsilon -\epsilon '}.
\end{equation}
Similarly, for the correlations  $\widetilde{\langle \sigma^z_i\sigma^z_j \rangle} $, we can write a measurement error matrix  
\begin{equation}
\begin{bmatrix} \tilde{P}_{\downarrow\downarrow} \\ \tilde{P}_{\uparrow\uparrow} \\ \tilde{P}_{\uparrow\downarrow / \downarrow\uparrow} \\
\end{bmatrix}
 =
  \begin{bmatrix}
    1-2\epsilon & \epsilon '^2 & \epsilon ' (1-\epsilon) \\
    \epsilon^2& 1-2\epsilon'&   (1-\epsilon ')\epsilon \\ 
    2\epsilon-\epsilon^2 & 2\epsilon'-\epsilon '^2 & 1-\epsilon-\epsilon'+2\epsilon\epsilon'
   \end{bmatrix}
   \begin{bmatrix} P_{\downarrow\downarrow} \\ P_{\uparrow\uparrow} \\ P_{\uparrow\downarrow / \downarrow\uparrow} \end{bmatrix},
\end{equation}
where $\tilde{P}_{\downarrow\downarrow}$ $(\tilde{P}_{\uparrow\uparrow}$) is the probability of measuring the two atoms in the ground-state (Rydberg state), and $\tilde{P}_{\uparrow\downarrow/\downarrow\uparrow}$ the probability to measure the atoms in different states. By inverting the error matrix, we get access to the real probabilities and $\langle \sigma^z_i\sigma^z_j \rangle = P_{\downarrow\downarrow}+P_{\uparrow\uparrow} - P_{\uparrow\downarrow / \downarrow\uparrow}$.

\subsection{Matrix-product-state emulation of interacting spins} For the emulation of the QPU dynamics we use the Pulser library~\cite{Silverio2022pulseropensource} and its backend EMU-MPS~\cite{Bidzhiev_pasqal_emulators_2025}, which exploits a matrix-product-state representation of the quantum state that is evolved in time by means of the time-dependent variational principle (TDVP) method. We use a maximum bond dimension of $\chi_\text{max}=200$ for $6\times 6$ systems. To compute the ground-state of the spin subsystem in the equilibrium loop we use the density-matrix renormalization group (DMRG) implemented in TeNPy~\cite{tenpy2024}, with also $\chi_\text{max}=200$.

\subsection{Emulation of noise} As the QPU is subject to various noise sources, including them in the numerical simulations is crucial for studying the compatibility of the QPU data with the theoretical model. 

In general, annealing protocols are expected to be more robust against stochastic noise sources because one follows adiabatically an eigenstate of the Hamiltonian. In accordance, we find that small discrepancies between the QPU annealing simulations and exact ground-state calculations (Fig.~\ref{fig:figure2} of the main text) seem very systematic and can be explained simply by small offsets in the QPU setpoint. In particular, we consider DMRG ground-state calculations at slightly different detuning setpoints $\delta \rightarrow \delta + \delta_0$, with $\delta_0=\pm 100$ KHz, which already explained the systematic offsets observed in the QPU data of Fig.~\ref{fig:figure2}. We expect the next corrections, due to Rabi frequency and positional offsets, as well as a finite annealing time, to have less impact than the detuning correction.

For the quench dynamics, stochastic noise sources are expected to play a bigger role, since one is interested in capturing coherent oscillations related to the population of different Hamiltonian eigenstates. Thus, besides residual systematic offsets in the laser setpoint $(\Omega, \delta$), we consider two main sources of stochastic noise for this scenario. On the one hand, shot-to-shot spatial noise in the atomic distances $r_{ij}$, coming from single-atom thermal motion with a position uncertainty of $\sigma_{xy}=0.2 \mu$m (in-plane) and $\sigma_{z}=0.8\,\mu$m (off-plane). On the other hand, laser phase noise~\cite{de_leseleuc_analysis_2018, kozlej_adiabatic_2025} leads to a time-dependent detuning noise on each shot of the form 

\begin{equation}
    \delta_{\text{LPN}}(t)=\sqrt{2}\sum_i \sqrt{(f_{i+1}-f_{i})S(f_i)}\cos\left[2\pi (f_it+\varphi_i)\right],
\end{equation}
where $S(f_i)$ is the power-spectral density in the relevant frequency support $\{f_i\}$ of the noise (measured experimentally), $t$ is the time of the simulation sequence, and $\varphi_i$ is a random phase that is sampled individually for each frequency following a flat probability distribution in the interval $[0, 2\pi)$. 

To numerically simulate these sources of noise we follow a classical Monte Carlo approach. First, we sample 50 instances of spatially noisy registers and laser phase noise sequences. Second, we use the same MPS numerical method described in the previous section to simulate each noisy instance. Finally, we average the observable of interest over the 50 noisy instances and compute from the histogram confidence intervals of $70\%$ around the mean value at each time, to extract the shaded regions of Fig.~\ref{fig:figure3}. Concerning the offsets, we include Rabi frequency offsets $\Omega_0/\Omega = (0.03, 0.04, 0.04)$ for the noisy emulation of Fig.~\ref{fig:figure3}(a), (b) and (c), respectively, according to the pre-calibrations performed before simulating the dynamics on the QPU. The detuning offsets were compensated in the simulated pulses.

\acknowledgments
We thank Christophe Jurczak for carefully reading the manuscript, and PASQAL Hardware and Cloud teams for their technical support. The work has been partly supported by EDF R\&D Advanced Simulation Program. Pasqal's team acknowledges funding from the European Union the projects PASQuanS2.1 (HORIZON-CL4-2022-QUANTUM02-SGA, Grant Agreement 101113690), shared with T. A., A. M., C. D. and J.M., and EQUALITY (Grant Agreement 101080142). This work (T.A.) is part of HQI initiative (www.hqi.fr) and is supported by France 2030 under the French National Research Agency award number “ANR- 22-PNCQ-0002".
This project (T.A.) has received funding from the European High-Performance Computing Joint Undertaking (JU) under grant agreement No. 101018180.

\section*{Supplemental Materials}
\setcounter{figure}{0}
\renewcommand{\thefigure}{S\arabic{figure}}
\subsection{Effect of the $H_s-H^\text{QPU}_s$ difference in equilibrium results}
To asses the impact of the approximation of the auxiliary spin Hamiltonian $H_s$ with $H_{QPU}$, here we show the expected difference in the equilibrium curves of the quasiparticle weight $Z_\text{bulk}$ for different system sizes. As explained in the Methods, the differences come from the $r^6$ long-range tail of interactions and edge $\sigma^z_i$ terms in the QPU Hamiltonian. 

\begin{figure}[h]
    \centering
    \includegraphics[width=1\linewidth]{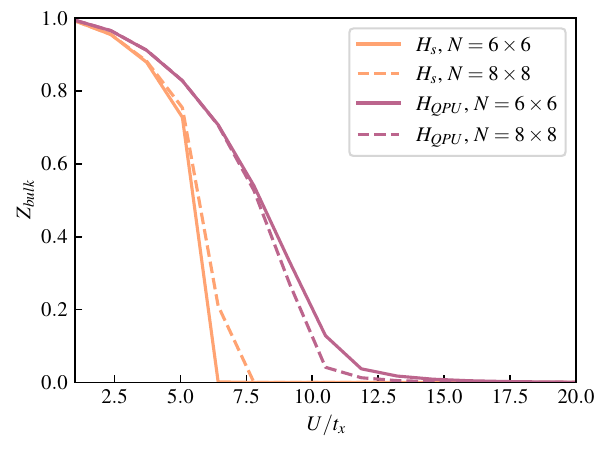}
    \caption{Discrepancy between the algorithm using the original auxiliary spin Hamiltonian $H_s$ and its QPU approximation $H_\text{QPU}$. The figure shows the evolution of the bulk quasiparticle weight obtained after 5 iterations loops at each $U/t_x$. Note that here we use the fermionic classical solver and the classical DMRG solver for both $H_s$ and $H_\text{QPU}$.}
    \label{fig:figureS1}
\end{figure}

For the case of the $r^6$ tail of $H_\text{QPU}$, while it is a fundamental difference between the two models, such fast-decaying interaction is only expected to lead to a slight shift of the interaction scale. Concerning the edge terms, they are not expected to affect the bulk physics in the large system size limit. We test this behaviour quantitatively by classically emulating the equilibrium auxiliary-spin self-consistent loop at $N=6\times 6$ and $N=8\times 8$, for the same model as in the main text ($t_y/t_x=0.65$). The results, shown in Fig.~\ref{fig:figureS1}, show that there is a tendency towards a reduced discrepancy between $H_{QPU}$ and $H_s$ for increasing system size. 

\subsection{Fourier spectroscopy of Hubbard quenches}
\begin{figure}[h]
    \centering
    \includegraphics[width=1\linewidth]{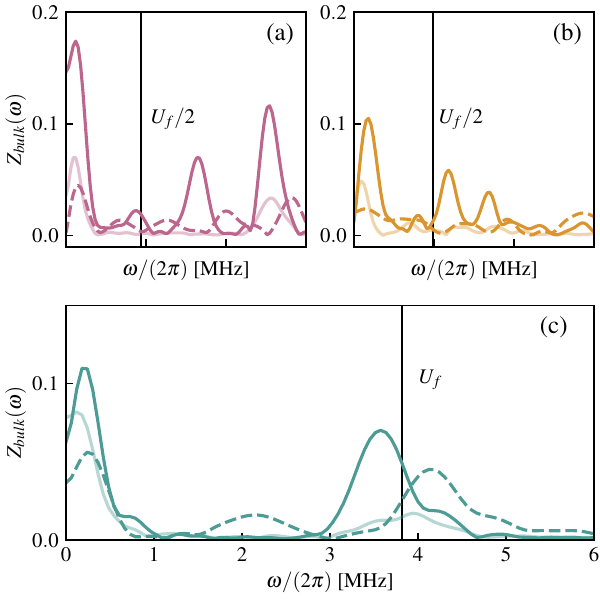}
    \caption{Frequency analysis of the quasiparticle weight oscillations from Fig.~\ref{fig:figure3}. The dashed line represents the smoothed QPU data.}
    \label{fig:figureS2}
\end{figure}

Here we show the time-Fourier transform of the quantities shown in Fig.~\ref{fig:figure3} of the main text. To compute this transform, we subtract the mean value of the signal in time, and apply a Gaussian time window to smoothen finite-time effects. The results are shown in Fig.~\ref{fig:figureS2}.

\vspace{2mm}
\FloatBarrier

% \bibliographystyle{apsrev4-2}
% \bibliography{bibliography} 


\end{document}